# Histopathological study on goldfish (*Carassius auratus*) gonads exposed to tobacco smoke

Parsa Khanghah A.[1*]

1-Department of Aquatic animal Health & Diseases, Faculty of Veterinay Medicine, Sanandaj Branch, Islamic Azad University, Sanandaj, Iran

*Corresponding author's Email: a.parsa@iausdj.ac.ir

**Abstract**

Ornamental fish have various positive effects in human life. Due to the effect and importance of aesthetics and artificial reproduction of these fish, which may be enclosed in aquarium environments and come in contact with cigarette smoke, the effects of tobacco smoke on the gonad tissue of goldfish were investigated. For this purpose, 60 goldfish randomly (weight 100 ± 3 gr) divided in 3 groups and were released in tanks containing 10 liters of water (temperature: 20 ± 2, hardness: 14 ppt, pH: 7/8). After adaptation, in treatment 1, 1gr of tobacco was heated daily with a direct flame, and the resulting smoke was collected and injected into water with an air pump, in treatment 2, this process was done twice a day. After 3 months, fish gonads tissue were sampled and histopathological slides were investigated. The results showed that in the treatment 2, there were early and immature oocytes in the ovarian tissue in comparison to other groups. Also, in the testes of fish of treatment 2, the reduction of spermatozoids and the higher number of spermatogonia were observed. In the treatment 3, these changes were more. A significant difference between the groups in both female and male was observed at the sexual maturation stages (P<0.001). Based on this study, the dissolution of tobacco smoke can have a negative effect on the process of sexual reproduction and fish exposed to more smoke are more likely to be sterile, and these changes were observed in both males and female.

**Introduction**

Ornamental fish are aesthetically superior to other fish. Breeding of these fish is important in terms of recreation, biodiversity, mental health, entertainment and tourist attraction (Vesal and Vosooghi, 2017). With increasing demand for ornamental fish, the production of these animals has become widespread in the industrial scale. Goldfish (*Carassius auratus*), as one of the mainly well-known ornamental fish, is a freshwater fish in the family Cyprinidae (Gumus *et al.*, 2016). It is the native of East Asia. The biological properties and low price of goldfish have led it to be used as a model for research (Blanco *et al.*, 2018).

One of the substances that affect the ornamental fish's environment is tobacco smoke, It is also common that in several places where these fish are kept, the presence of people who use tobacco is noticeable) Also, it should be noted that the tobacco can be used to fight against parasitic infections of fish (Woo, 1997). Tobacco smoke contains over 4000 components including nicotine, aldehydes, alkaloids, polycyclic aromatic hydrocarbons, hydrogenated compounds, radioactive substances, heavy metals, etc (Dai *et al.*, 2015). These substances produce numerous biological effects in the body tissues. Human studies have shown the detrimental impact of tobacco smoke on male fertility (Kovac *et al.*, 2015).

On the one hand, due to the exposure of the ornamental fish's environment to tobacco smoke, and on the other hand, the lack of pathological studies on the effect of tobacco smoke on ornamental fish gonads, the study design is necessary in this regard. Hence, the aim of this study was to evaluate the effect of tobacco smoke injection in water on histopathological changes of goldfish gonads.

**Materials and methods**

This experimental study was carried out in the nine aquariums (Dimensions 31 × 48 × 28 cm) with a capacity of 10 liters under controlled environmental conditions (Chlorine-free water temperature of 20 ± 2, 14/10 h light/dark cycle, water hardness of 14 ppt, pH = 7/8, and proper aeration and nutrition). The study began 7 days after fish adaptation to environmental conditions. In this study, the use of animals was based on international ethical rules (Clark *et al.*, 1997).

*Study design*

A total of 30 apparently healthy goldfish (*Carassius auratus*) with average weight 100 ± 3 gr were randomly selected and divided into three equal groups (10 fish per tank) as follows:

Control group (C): Healthy goldfish ( no contact with tobacco smoke)

Treatment 1 (T1): Healthy goldfish + tobacco smoke (once a day until 1gr tobacco is completely burnt);

Treatment 2 (T2): Healthy goldfish + tobacco smoke (twice a day until 1gr tobacco is completely burnt).

For each group, three replications were considered. In order to inject tobacco smoke into the water, 1 gr of tobacco was placed in a metal basket. In the next, it was exposed under the heat of the flame. Finally, the smoke from the burning tobacco was completely injected into the water through a funnel with a glass hose attached to the aeration sucker. The study period was 3 months for all groups.

*Behavioral pattern assessment*

Behavioral patterns such as the erratic movement, leaping and instabilities of fish were examined during the study and no abnormality was observed.

*Sampling*

At the end of the study, fish were anesthetized by clove powder and they were euthanized. Fish abdominal cavity was incised and gonads tissue collected and kept in the 10% formalin buffer fixative solution.

*Tissue preparation*

After 48 hours of fixation, the samples were embedded in paraffin. Then, paraffin blocks were cut by a microtome (5-μm) and stained with hematoxylin and eosin (H&E).

*Histopathological evaluations*

Each section was examined by two pathologists (double-blind) with an optical microscope. Accordingly, the frequency of sexual maturation was determined in the groups. Slides were also assessed for pathological lesions such as necrosis, hyperemia and leukocyte infiltration.

*Statistical analysis*

Data were analyzed by Chi-square test. The statistical analyses were performed by GraphPad Prism (Version 7.03) software (GraphPad Software, San Diego, California, USA).

This research is approved by the ethics committee of the Islamic Azad University

This research continued for 4 months and all sections of it were done in the faculty of Veterinary Medicine, Islamic Azad University, Iran

**Results**

There were no visible changes in fish behavior patterns. In this study, the significant difference between the groups in both female and male genders was observed at gamete sexual maturation ($P < 0.001$). After 3 months, as shown in Fig. 1A, the majority of the control fish were in stages 4 and 5 of sexual maturation. Furthermore, most of the fish in the T2 and T3 groups were at stages 2 and 3 of sexual maturity (Fig.1).

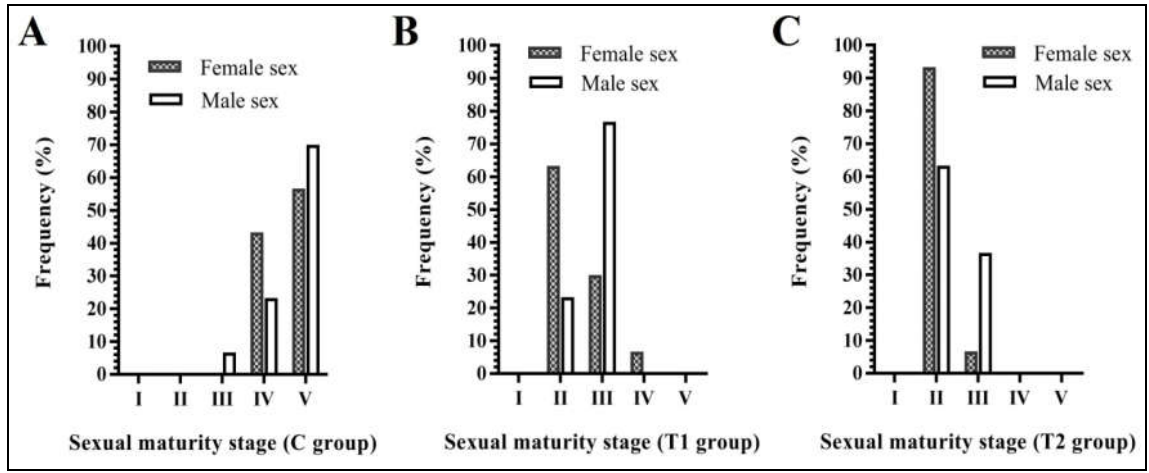

Figure 1: The frequency of different stages of sexual maturation in the groups, 3 months after the initiate of the study. (A)=Control-(B)=Treatment 1. (C)=Treatment 2.

Pathological lesions (e.g., necrosis, hyperemia and leukocyte infiltration) were not observed. The results of histopathology and the various stages of maturation are shown (Figs. 2 and 3).

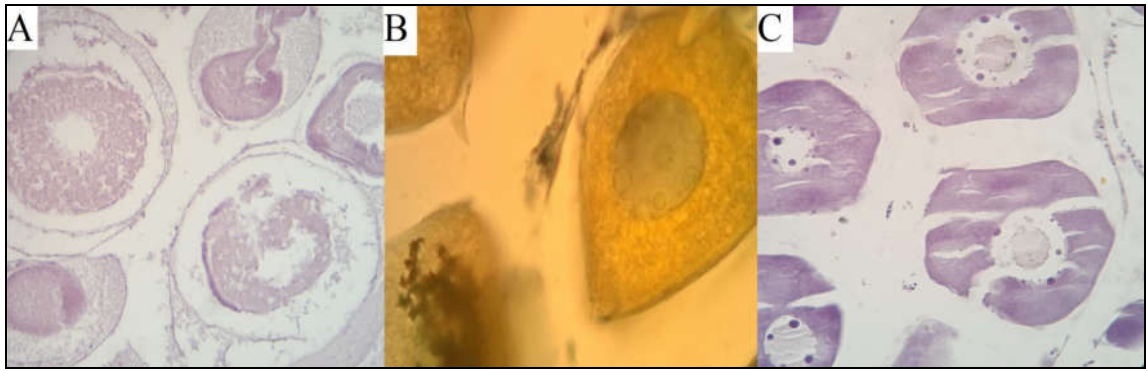

Figure 2: Histopathological findings of ovarian tissue in the groups (H&E). (A) Ovarian tissue of the control group, the majority of oocytes were in the fifth stage of sexual maturation and ovulation is observed (x40). (B) In the ovarian tissue of the T1, the majority of oocytes were in the second stage of sexual maturation and nucleoli arrangement on the periphery (x100). (C) In the ovarian tissue of the T2, the majority of oocytes are in the second stage of sexual maturation and nucleoli arrangements are seen on the periphery (x40).

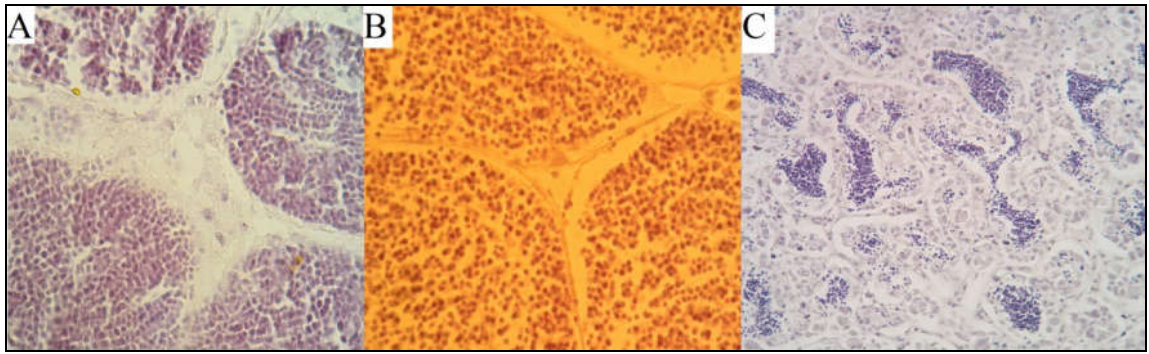

Figure 3: Histopathological findings of testicular tissue in the groups (H&E). (A) Testicular tissue of the control group, spermatozoids were observed in the vas deferens (x100). (B) Testicular tissue of the treatment 1, stage 3 has been observed and spermatozoids have not been made (x100). (C) Testicular tissue of the treatment 2, stage 3 has been observed and spermatozoids have not been made (x40).

**Discussion**

Few studies have been conducted on the effects of tobacco smoke on the growth and reproduction of fish. Most of these studies were carried out on the zebrafish, as a bridge model. In line with this study, in a study, cigarette smoke leads to delayed growth of zebrafish larvae, as well as the death of most embryos (Hammer *et al.*, 2011). In another study, which aims to create a model for cigarette smoke toxicity in the zebrafish larvae, researchers reported that high concentrations of cigarette smoke (over 20 μg/ml) could reduce development, increase neurotoxicity and reduce light and dark response levels, and at low concentrations (10 and 15 μg/ml of cigarette smoke) lead to improved growth, decreased neurotoxicity and increased light and dark response levels (Ellis *et al.*, 2014).

Contrary to our study, in an evaluation of the daily exposure to smoke of commercial cigarettes on vital organs of *Carassius auratus*, the pathological lesions were shown in the swim bladder (degeneration with severe fibrosis), liver (the hypertrophy of hepatocytes, dilation of sinusoids, and focal necrosis) and blood vessels (arterial blockage) (Kupekar and Patil, 2015).

Divergent to the results of behavioral patterns in the present study, the effect of tobacco (*Nicotiana tobaccum*) leaf dust on the African catfish (*Clarias gariepinus*) showed changes in behavioral patterns such as erratic swimming, uncoordinated movement, and respiratory distress (Kori-Siakpere and Oviroh, 2011).

Nicotine is one of the most important components of tobacco smoke (Dai *et al.*, 2015). In a study, increasing nicotine concentration (5 to 40 μM) causes mortality and abnormalities in the heart, as well as increasing apoptotic cells in neuromasts of the zebrafish (Yoo *et al.*, 2018). In addition, other studies have been conducted to investigate the effects of tobacco smoke in fish demonstrated that delayed growth, abnormalities in the head and eyes, interrupted tail, heart edema, body deviation, brain hemorrhage, and angiogenesis in the head area (Massarsky *et al.*, 2015; Parker and Connaughton, 2007). The researchers found the chemicals from just one filtered cigarette butt had the ability to kill fish living in a one-liter bucket of water and nicotine is toxic to microbes, plants, benthic organisms, bivalves, zooplankton, fish and mammals, then the effect of tobacco smoke in this research on the gonads of fish, which are simple vertebrate, can be probability (Slaughter *et al.,* 2011).

The results of another study are similar to the results of this study, exposure particulate phase of cigarette smoke leads to long-lasting sex-specific behavioral outcomes in the adult zebra fish (Massarsky *et al.,* 2018).

The results of the current study showed that tobacco smoke can play a significant role in reducing the function of the gonads and delaying sexual maturation stages. Consequently, with this finding, certain breeding objectives can be followed, for example, this finding may be used in the sterilization of aquarium fish. It should be noted that supplementary studies are necessary in this regard.

**Acknowledgments**

We thank the staff of the Faculty of Veterinary Medicine, Islamic Azad University, Sanandaj